\documentclass[aps,pre,reprint,showpacs,nofootinbib]{revtex4-1}

\usepackage{graphicx}
\usepackage{newtxtext,newtxmath}
\usepackage{amsmath}
\usepackage{upgreek}
\usepackage{units}
\usepackage{xspace}

\usepackage[bookmarks,colorlinks,breaklinks]{hyperref}
\hypersetup{linkcolor=blue,citecolor=blue,filecolor=black,urlcolor=blue}

\newcommand{\mum}{\upmu\mathrm{m}}
\newcommand{\ccm}{cm^{-3}}
\newcommand{\Tieff}{\ensuremath{T_i^\textrm{eff}}\xspace}
\newcommand{\dif}[1]{\mathrm{d}#1}

\begin{document}

\title{Turbulent stagnation in a z-pinch plasma}

\author{E. Kroupp}
\author{E. Stambulchik}
\email{Corresponding author: evgeny.stambulchik@weizmann.ac.il}
\author{A. Starobinets}
\author{D. Osin}
\altaffiliation[Present address: ]
   {Tri Alpha Energy Inc, Foothill Ranch, CA, USA}
\author{V. I. Fisher}
\author{D. Alumot}
\altaffiliation[Present address: ]
   {Applied Materials, Rehovot, Israel}
\author{Y. Maron}
\affiliation{Faculty of Physics, Weizmann Institute of Science, Rehovot 7610001,
             Israel}
\author{S. Davidovits}
\email{Corresponding author: sdavidov@princeton.edu}
\author{N. J. Fisch}
\affiliation{Princeton University, Princeton, New Jersey 08540, USA}
\author{A. Fruchtman}
\affiliation{H.I.T.---Holon Institute of Technology, Holon 5810201, Israel}


\begin{abstract}
The ion kinetic energy in a stagnating plasma was previously determined by
Kroupp \emph {et~al.}, \BibitemOpen \href {\doibase 10.1103/PhysRevLett.107.105001}
{\bibfield {journal} {\bibinfo  {journal} {Phys. Rev. Lett.}\ }\textbf {\bibinfo
{volume} {107}},\ \bibinfo {pages} {105001} (\bibinfo {year}
{2011})}\BibitemShut {NoStop}, from Doppler-dominated lineshapes augmented by
measurements of plasma properties and assuming a uniform-plasma model. Notably,
the energy was found to be dominantly stored in hydrodynamic flow. Here, we
advance a new description of this stagnation as supersonically turbulent. Such
turbulence implies a non-uniform density distribution. We demonstrate how to
re-analyze the spectroscopic data consistent with the turbulent picture, and
show this leads to better concordance of the overconstrained spectroscopic
measurements, while also substantially lowering the inferred mean density.
\end{abstract}



\maketitle

\textbf{Introduction} ---
In implosions of a cylindrical z-pinch plasma, hydrodynamic kinetic
motion is ultimately transferred to thermal motion of plasma
particles---electrons and ions---through a cascade of atomic and thermodynamic
processes~\cite{ryutov:2000a,slutz:2012a, herrmann:2014a, sinars:2016a}. These
processes culminate at the stagnation phase, producing high-energy-density
plasmas and generating powerful x-ray and neutron
radiation~\cite{giuliani:2016a}. 

Previous x-ray spectroscopic analysis of pinch plasmas~\cite{kroupp:2011a,
maron:2013a} found that the ion kinetic energy at the stagnation phase was
dominantly non-thermal hydrodynamic motion, while the plasma appeared largely
uniform at spatial and temporal scales down to at least $\unit[100]{\mum}$ and
$\sim$\unit[1]{ns}, respectively. An explanation of this phenomenon has been
offered~\cite{giuliani:2014a}, where simulations showed steep radial velocity
gradients in the stagnation region. On the other hand, we note here that the
Reynolds number in the stagnating plasma is initially high ($\sim\!10^5$),
making turbulence a candidate for such significant small-scale hydrodynamic
motion (the 2D simulations in~\cite{giuliani:2014a} would not be expected to
reproduce turbulent behavior). The inferred Mach numbers, $M$, at stagnation are
supersonic, see Table~\ref{tbl:data} for Re and $M$. Were supersonic turbulence
present, it would imply substantial nonuniformity in quantities such as the
density (see, e.g., Fig.~2 in \citep{federrath:2010a}). However, the previous
analysis~\cite{kroupp:2011a} assumed a uniform plasma.

This study re-analyzes the experimental data~\cite{kroupp:2011a} without
assuming uniformity, using, instead, a modeled turbulent density
distribution~\cite{hopkins:2013a}. In doing so, we give both a new physical
description of a stagnated plasma dominated by supersonic turbulence, and a new
spectroscopic analysis method. We find full (actually, improved) consistency
with the observations, but a significantly (about two-fold) lower average
density. The results are believed to also be relevant for large-scale z-pinch
devices, inertial confinement experiments with large residual hydrodynamic
motion, and in various astrophysical contexts.

\textbf{Short description of the previous study} --- 
In brief, a 9-mm-long neon-puff z-pinch was imploded in \unit[500]{ns} under a
current rising to \unit[500]{kA} at the stagnation time. Experimental
diagnostics included high-resolution ($\sim$$\unit[200]{\mum}$) gated x-ray
filtered-pinhole imaging, a spectrometer recording Ne He-like dielectronic
satellites with a resolving power of 6700, and a photo-conductive detector (PCD)
sensitive to $\hbar\omega \gtrsim \unit[700]{eV}$ radiation. All the data were
{\em simultaneously} acquired over the stagnation period, about $\unit[\pm
5]{ns}$ around the peak of the PCD signal with a time resolution of
$\sim$\unit[1]{ns}. A plasma segment at $z = \unit[5 \pm 1]{mm}$ along the
pinch axis was used for the analysis.

The modeling assumed a uniform-cylinder plasma with a prescribed (within
experimental uncertainties) time evolution of $T_e$, $n_e$, and plasma radius
$r_\mathrm{pl}$. The experimental data and uniform model parameters are shown in
Table~\ref{tbl:data}. Assuming uniformity, the electron density $n_e$ was
determined based on the satellite-intensity ratio~\cite{seely:1979a,
kroupp:2007a}; it is $n_e^0$ in the table. A separately measured time-integrated
continuum slope~\cite{alumot:2007a} was found to agree with the $T_e(t)$
assumed. The x-ray images give $r_{\rm{pl}}$ to within the
$\sim$$\unit[200]{\mum}$ resolution, $r_\mathrm{min},r_\mathrm{max}$ in
Table~\ref{tbl:data}. The self-consistency of the uniform-model time
dependencies for $n_e$, $T_e$ and $r_\mathrm{pl}$ was verified using the
additional measurement of the absolutely calibrated PCD signal, which is
sensitive to all three quantities. With $T_e$ fixed, it was found that
\emph{either} $n_e$ \emph{or} $r_{\rm{pl}}$ could be taken at the center of its
measured value range (i.e. uncertainty), with the other quantity then within one
or two standard deviations of its independently measured value. With $n_e$ the
more important quantity, it was chosen to let $r_{\rm{pl}}$ vary outside one
deviation; this gave $r_\mathrm{pl}^0$ in Table~\ref{tbl:data}.

\textbf{New model} --- 
Although the uniform plasma analysis was reasonably consistent with the data,
the uniform density assumption will not be \emph{physically sound} if the plasma
is highly turbulent, as expected for the measured \rm{Re} and $M$. Therefore, we
use a (non-uniform) turbulent plasma model. Within such a model, all plasma
properties---density $\rho$, electron $T_e$ and ion $T_i$ temperatures, and the
non-thermal ion velocity $v_\mathrm{flow}$---have certain distributions, with
possible correlations between them. This work analyzes the simplest case, of
isothermal turbulence, where $T_e$ and $T_i$ are uniform (see the discussion
below), whereas $\rho$ (and $n_e$) are not. Then, the previous spectral fits and
$T_e$ analysis are still valid because: the correlation between turbulent
velocity and density is very weak for isothermal
turbulence~\cite{federrath:2010a}; and the turbulent velocity distribution is
well approximated by a Gaussian (see, e.g.~\cite{smith:2000a, porter:2002a,
kitsionas:2009a}), which was also the assumed non-thermal velocity distribution
used in the original study~\cite{kroupp:2011a}. This leaves \Tieff inferred from
Doppler broadening~\cite{kroupp:2007a} unaffected. We now show that, using a
turbulent density probability distribution function (PDF) that is consistent
with the measured $\rm{Re}$ and $M$, the inferred density is substantially
reduced, which allows the inferred plasma radius at each time to be larger,
while staying consistent with the PCD signal. This larger radius now agrees well
with $\left[r_{\rm{min}},r_{\rm{max}}\right]$ (except at the first measurement
time, which has the weakest signal). Thus, the new turbulent model is an
improvement \emph{both} because it is physically sound \emph{and} gives an
improved match to the observations.

We work with electron density $n_e$ instead of mass density $\rho$, since the
atomic experimental data are sensitive to $n_e$. The two are related, $\rho =
\langle Z_i \rangle^{-1} m_i n_e$, where $\langle Z_i \rangle$ is the mean ion
charge and $m_i$ is the ion mass. In principle, $\langle Z_i \rangle$ is a
function of $T_e$ and $n_e$, but for the ranges of plasma parameters of
interest, it varies very weakly~\cite{kroupp:2011a, giuliani:2014a}, so we
assume  $\rho \propto n_e$.

For each measurement the density has a PDF, $P(n_e)$. The previous
data analysis~\cite{kroupp:2011a} corresponds to $P(n_e) \equiv \delta(n_e -
n_e^0)$. $P(n_e)$'s are different at different times and
$z$-positions, i.e., $P(t, z; n_e)$; for brevity, these $t, z$ labels will be
omitted.

Let us switch to dimensionless quantity
\begin{equation}
    \label{eq:PDFnormalization}
    \xi \equiv n_e/n_e^0; \int P(\xi)\, \dif{\xi} = 1 .
\end{equation}
The average density is $\langle n_e \rangle = n_e^0 \int \xi\, P(\xi)\,
\dif{\xi}$. It is important to note that $\langle n_e \rangle$ is not the same
as $n_e^0$. The nonuniform density affects two of the previous measurements:
$n_e$ from line ratios and the absolutely calibrated PCD signal, from which one
can infer the radiating mass (product of $n_e$ and $r_{\rm{pl}}^2$), for a given
$T_e$. These measurements give two constraints on the turbulent PDF, $P(\xi)$,
which thus determine the new mean density.

Assuming the collisional-radiative equilibrium is established much faster than
the hydromotion, the intensity of a discrete spectral line
or continuum radiation in a turbulent plasma can be obtained in the static
approximation~\cite{stamm:2017a}, viz.,
\begin{equation}
    \label{eq:I}
    \langle I \rangle = \int \alpha(\vec{r}) d^3 r =
    \pi r_\mathrm{pl}^2 \ell
        \int \alpha(\xi) P(\xi)\, \dif{\xi} .
\end{equation}
Here, $\alpha$ is the local plasma emissivity, approximately scaling as $\propto
\xi^2$ if the density does not vary too much, $\ell$ is the length (in the $z$
direction) of the plasma segment being analyzed, and we assumed that density
variations are independent of $r$.
In particular, the PCD signal is
\begin{equation}
  \label{eq:I_PCD}
  I_\mathrm{PCD} \propto
 \pi r_\mathrm{pl}^2 \ell \int \xi^2 P(\xi)\, \dif{\xi}.
\end{equation}
 Using this, and the fact that the previous model described $I_\mathrm{PCD}$
self-consistently (within the errors bars $\delta I_\mathrm{PCD}$) by assuming
$r_\mathrm{pl}^0$, we can get a first constraint on $P(\xi)$, bounding
$I_\mathrm{PCD}$ with $r_\mathrm{min}$ and $r_\mathrm{max}$ and $\pm \delta
I_\mathrm{PCD}$,
\begin{equation}
    \label{eq:constraint2}
    \left(1 - \frac{\delta I_\mathrm{PCD}}{I_\mathrm{PCD}}\right)
    \left(\frac{r_\mathrm{pl}^0}{r_\mathrm{max}} \right)^2 \leq
    \int \!\! \xi^2\, P(\xi)\, \dif{\xi} \leq 
    \left(1 + \frac{\delta I_\mathrm{PCD}}{I_\mathrm{PCD}}\right)
    \left(\frac{r_\mathrm{pl}^0}{r_\mathrm{min}} \right)^2 \,.
\end{equation}

\begin{figure}
  \includegraphics[width=\linewidth]{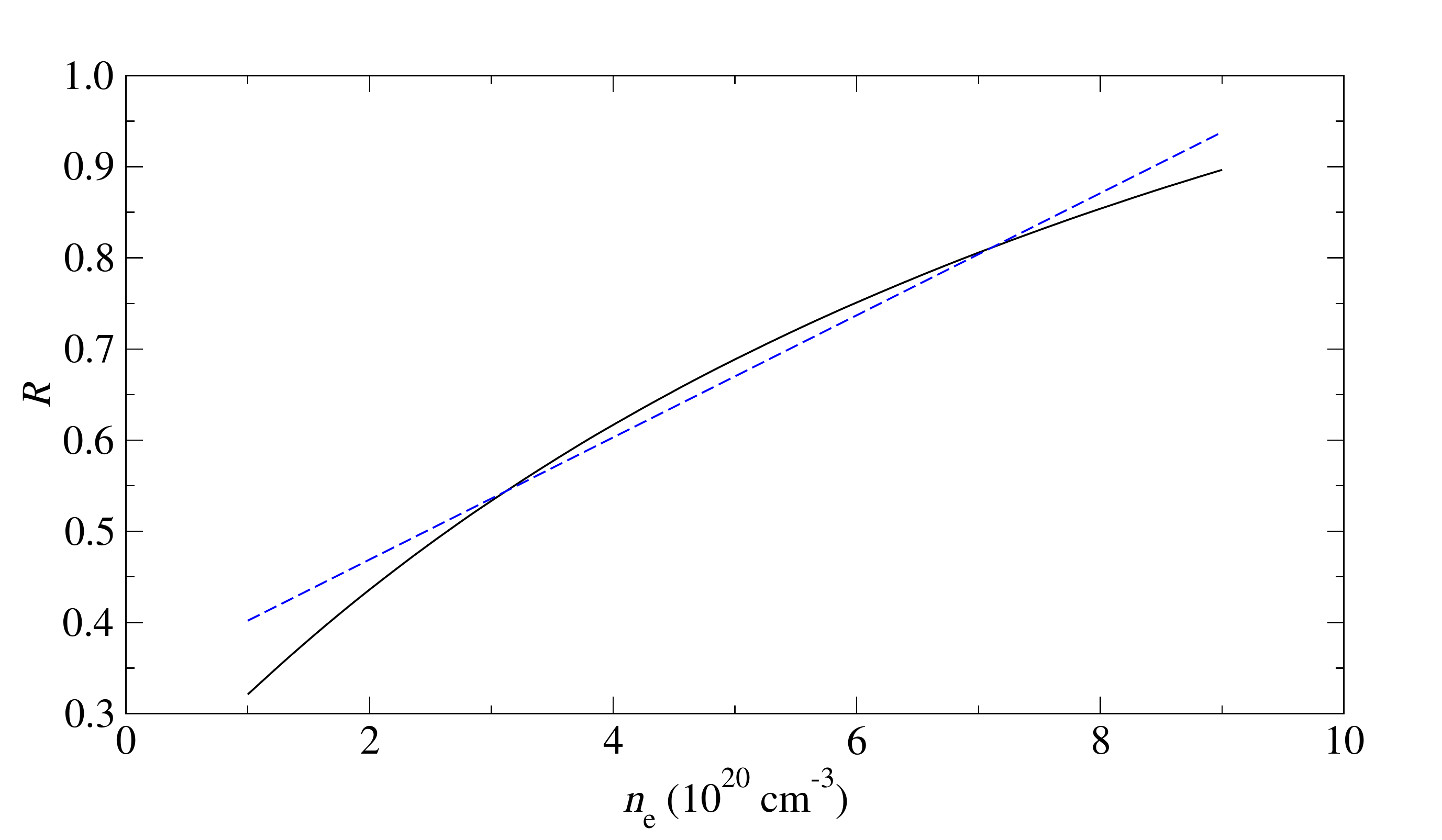}
  \caption{$2p^2 (^3P) \rightarrow 1s2p(^3P)$ to $2s2p (^3P) \rightarrow
           1s2s(^3S)$ intensity ratio (the solid line) and its linear
           approximation (the dashed line)
           calculated~\cite{ralchenko:2001a} as a function of $n_e$.
           Optically thin steady-state plasma with $T_e =
           \unit[200]{eV}$ is assumed.}
  \label{fig:R}
\end{figure}

Some of the autoionizing dielectronic satellites have even
stronger density dependence than $\propto \xi^2$---which is why the intensity
ratio of such a satellite to another line (in our case---another close-by
dielectronic satellite) allows for inferring the density~\cite{seely:1979a}.
Both dependencies are complex, but around the density point of interest
($\sim \unit[5\times10^{20}]{\ccm}$), their ratio is rather close to a linear
form, $R \approx R^0 + a_R(n_e/n_e^0 - 1)$, in a steady-state optically thin
plasma (see Fig.~\ref{fig:R}). Hence, if $n_e$ does not vary too wildly,
(say, within a factor $\times 2$ in each direction), 
\begin{equation}
    \label{eq:R}
    \langle R \rangle = R^0 +
        a_R \frac{\int (\xi - 1) \xi^2\, P(\xi)\, \dif{\xi}}
                 {\int \xi^2\, P(\xi)\, \dif{\xi}} .
\end{equation}

The measured quantity $R_\mathrm{expt}$ is known within its error bars, i.e., 
$\langle R \rangle = R_\mathrm{expt} = R^0 \pm \delta R$. Therefore,
Eq.~(\ref{eq:R}) gives a second constraint  on $P(\xi)$,
\begin{equation}
    \label{eq:constraint1}
    1 - \frac{\delta R}{a_R} \le
    \frac{\int \xi^3
             P(\xi)\, \dif{\xi}}
             {\int \xi^2
             P(\xi)\, \dif{\xi}}
    \le 1 + \frac{\delta R}{a_R} \,.
\end{equation}

To model the density PDF that would result from turbulence in the stagnating
plasma, we use the PDF of \citet{hopkins:2013a}.
Since the model assumes the average density is known, it is convenient to
introduce dimensionless {\em volumetric} density by normalizing to $\langle n_e
\rangle$, i.e., $\xi_V \equiv n_e/\langle n_e \rangle$. Evidently, $\xi/\xi_V =
\langle n_e \rangle/n_e^0$. In terms of $\xi_V$, the (volumetric) PDF is
\begin{align}
    P_V \! \left(\xi_V \right) \rm{d} \xi_V =\, &
    \frac{I_1\!\left(2 \sqrt{\lambda \omega(\xi_V)} \right)}
         {\exp\!\left[ \lambda + \omega(\xi_V) \right]}
    \sqrt{\frac{\lambda}{\theta^2 \omega(\xi_V)}}
    \frac{\rm{d} \xi_V}{\xi_V},
    \label{eq:hopkinsPDF}
\end{align}
where $\lambda \equiv \sigma_{s,V}^2/2 \theta^2$, and $\omega (\xi_V) \equiv
\lambda/(1+\theta) - \rm{ln} (\xi_v)/\theta$, and $I_1$ is the modified
Bessel function of the first kind. This two-parameter PDF depends on
a variance, $\sigma_{s,V}^2$, and a measure of intermittency, $\theta$. As
$\theta \rightarrow 0$, the PDF becomes lognormal. This PDF fits well for
simulations conducted at a wide range of Mach
numbers~\cite{hopkins:2013a}. Although we presently treat the turbulence as
isothermal, this PDF has been shown to fit for simulations of non-isothermal
turbulence~\cite{federrath:2015a}. In general, the values of, $\sigma_{s,V}^2$,
$\theta$, depend on the turbulence properties; they are typically modeled as
depending on the turbulent Mach number, the mix of compressive and solenoidal
forcing, and, in the non-isothermal case, the polytropic
gamma~\cite{hopkins:2013a,federrath:2015a}. As such, the turbulence model
does not introduce any ``free'' parameters, since its parameters vary only as a
direct consequence of  the variation of measured or inferred plasma properties.

For the value of $\theta$, we use the fit to simulation
data~\cite{hopkins:2013a}, which is $\theta \approx 0.05 M_c$. Here $M_c$
is the compressive Mach number, also written $M_c = b M$~\cite{konstandin:2012a,
konstandin:2016a}, and $b$ is related to the mix of solenoidal and compressive
modes~\cite{federrath:2008a, konstandin:2012a, konstandin:2016a}.
For the density variance, $\sigma_{s,V}^2$, we combine the usual isothermal
logarithmic density variance (see, e.g.,~\citep{padoan:1997a, passot:1998a,
padoan:2011a, molina:2012a}), $\sigma_s^2 \approx \ln \! \left[ 1+ b^2 M^2
\right]$, with the relationships $\sigma_{s,V}^2 = (1 + \theta)^3
\sigma_{s,M}^2$~\citep{hopkins:2013a} and $\sigma_s^2 = \sigma_{s,V}
\sigma_{s,M}$~\citep{federrath:2015a}. This yields $\sigma_{s,V}^2 = \left( 1 +
\theta \right)^{3/2} \ln \! \left[1 + b^2 M^2 \right]$.
Here we take $b = 0.4$; see the discussion below for more on this choice, and
caveats associated with the turbulence model.

The Mach number at each time is calculated using the data in Table
\ref{tbl:data}; $M = v_\mathrm{flow}/c_s$, where
$v_\mathrm{flow} = \left[3 (\Tieff - T_i)/m_i \right]^{1/2}$
and $c_s = \left[ \gamma (T_e n_e + T_i n_i)/ \left(n_i m_i + n_e m_e\right)
\right]^{1/2} $, where $\gamma = 1$ is used, assuming isothermality (discussed
below).

\begin{figure}
  \includegraphics[width=\linewidth]{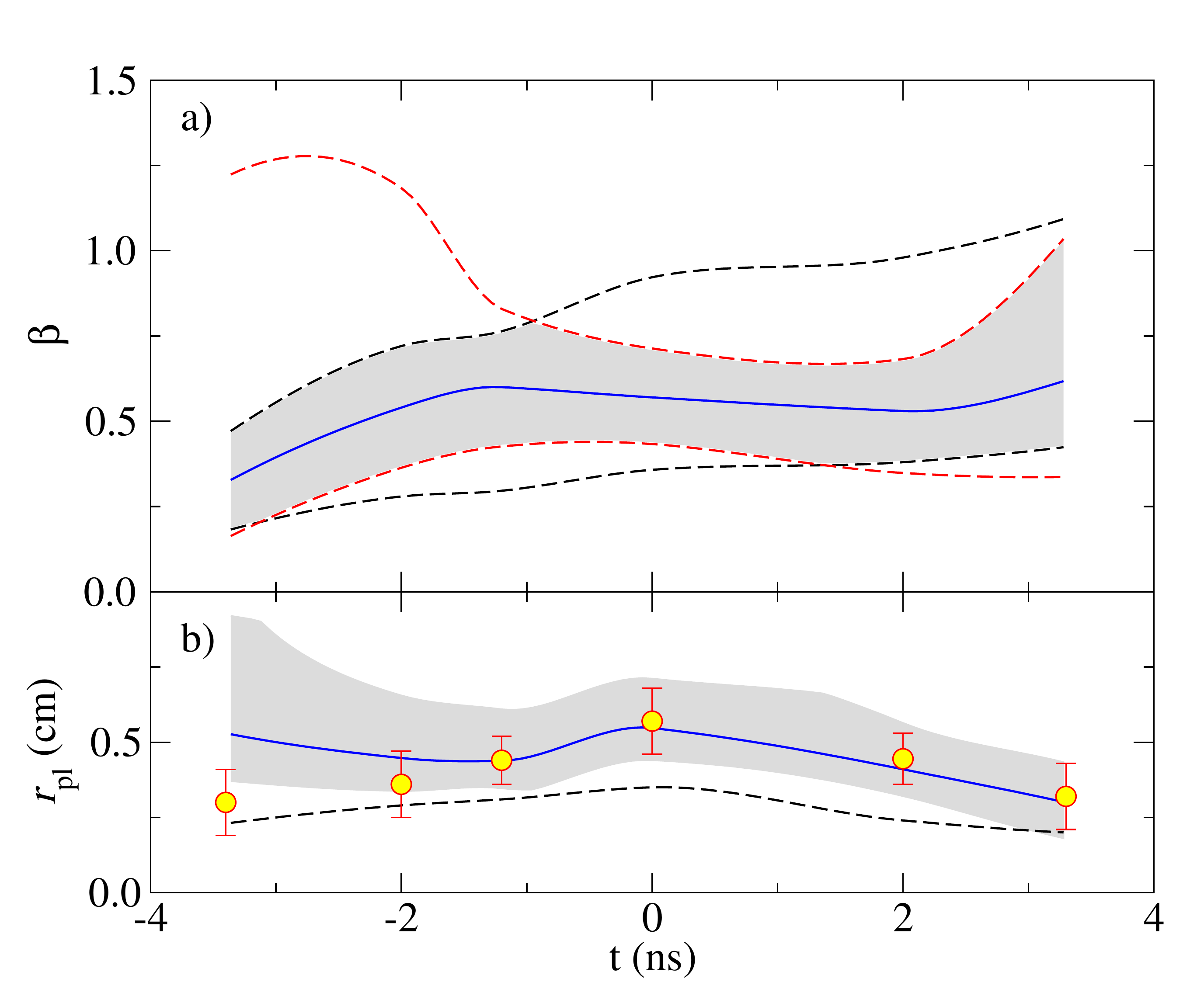}
  \caption{a) Limits of the double inequalities (\ref{eq:beta2}) and
           (\ref{eq:beta1}) are shown as red (gray) and black dashed lines,
           respectively. The ranges of $\beta$ (corrected density
           $n_e^\mathrm{turb} = \beta n_e^0 $)  satisfying both inequalities are
           designated by the gray filled area, with the tentative values used to
           correct the uniform-model parameters indicated by the solid line. b)
           $r_\mathrm{pl}^\mathrm{turb}$ (the solid line, with the grey area
           denoting uncertainties) shows an improved agreement with the
           experimental data (symbols with error bars). $r_\mathrm{pl}^0$ of the
           uniform-plasma model is given by the dashed line.}
  \label{fig:beta}
\end{figure}

\textbf{Results and discussion} ---
We now use the turbulent density PDF, Eq.~(\ref{eq:hopkinsPDF}), in the
constraints~(\ref{eq:constraint2}) and (\ref{eq:constraint1}). In addition to
satisfying the usual normalization condition, Eq.~(\ref{eq:PDFnormalization}),
it also conserves the average density, $\int \xi_V P_V \! (\xi_V) \rm{d} \xi_V =
1$. However, experimentally the average density is unknown; in order to use the
volumetric PDF and its moments, we connect $\xi$ and $\xi_V$ with a free
parameter $\beta$, $\xi = \beta \xi_V$. Once the turbulence PDF satisfying the
experimental data within the constraints (\ref{eq:constraint2}) and
(\ref{eq:constraint1}) is determined, $\beta n_e^0$ will give the new mean
density, corrected for the presence of turbulence; more generally,
$\langle\xi^k\rangle = \beta^k \langle\xi_V^k\rangle$. With this in mind,
Eqs.~(\ref{eq:constraint2}) and (\ref{eq:constraint1}) become a set of
inequalities on $\beta$,
\begin{align}
    \label{eq:beta2}
    \sqrt{\frac{1 - \frac{\delta I_\mathrm{PCD}}{I_\mathrm{PCD}}}
    {\langle\xi_V^2\rangle}}
    \frac{r_\mathrm{pl}^0}{r_\mathrm{max}}
    \leq \beta \leq 
    \sqrt{\frac{1 + \frac{\delta I_\mathrm{PCD}}{I_\mathrm{PCD}}}
    {\langle\xi_V^2\rangle}}
    \frac{r_\mathrm{pl}^0}{r_\mathrm{min}} \, \\
    \label{eq:beta1}
    \left(1 - \frac{\delta R}{a_R}\right)
    \frac{\langle\xi_V^2\rangle}{\langle\xi_V^3\rangle} \le
    \beta 
    \le \left(1 + \frac{\delta R}{a_R}\right)
    \frac{\langle\xi_V^2\rangle}{\langle\xi_V^3\rangle}\,,
\end{align}
shown graphically in Fig.~\ref{fig:beta}a. The new model predicts a
significantly (about two-fold) lower average density. With $\beta$ chosen, the
plasma radius needs to be corrected, accounting for the turbulence-modified
average emissivity. Using Eq.~(\ref{eq:I_PCD}), it follows that
$r_\mathrm{pl}^\mathrm{turb} = r_\mathrm{pl}^0/\sqrt{\langle\xi^2\rangle} =
r_\mathrm{pl}^0/(\beta \sqrt{\langle\xi_V^2\rangle})$. Notably,
$r_\mathrm{pl}$'s in the present model (listed as $r_\mathrm{pl}^\mathrm{turb}$
in Table~\ref{tbl:data}) fit the measured values better than the original
model~\cite{kroupp:2011a}, as shown in Fig.~\ref{fig:beta}b.

\begin{table*}
    \caption{\label{tbl:data}The experimental data~\cite{kroupp:2011a} relevant
             for the analysis presented; the plasma parameters assumed for
             ($r_\mathrm{pl}^0$, $n_e^0$, $T_e$) and inferred from ($T_i$,
             $M$, Re)
             the {\em uniform}-plasma modeling; the calculated isothermal
             turbulence parameters, volumetric density factor $\beta$ and
             respectively corrected plasma electron density and radius. Units
             are as follows: all radii are in mm, all temperatures are in eV,
             and densities are in $\unit[10^{20}]{\ccm}$.}
    \begin{tabular}{cccccc|cccccc|ccccccc}
    \hline
    \multicolumn{6}{c|}{Experimental data} &
    \multicolumn{6}{c|}{Uniform plasma} &
    \multicolumn{7}{c}{Isothermal turbulence}\\
    t (ns) & $\delta R$ & $I_\mathrm{PCD}$ (GW) &
    $r_\mathrm{min}$ & $r_\mathrm{max}$
    & \Tieff  
    & $r_\mathrm{pl}^0$ & $n_e^0$ & $T_e$  &
    $T_i$ & $M$ & Re &
    $\theta$  & $\sigma_{s,V}^2$ 
    & $\langle \xi_V^2 \rangle$ & $\langle \xi_V^3\rangle$ &
    $\beta$ & $n_e^\mathrm{turb}$ & $r_\mathrm{pl}^\mathrm{turb}$ \\
    \hline
    -3.4 & 0.15 & $0.35 \pm 0.3$ & 0.19 & 0.41 & 3000 & 0.23 & 6.0 & 120 & 250 & 2.4 & $8.1\times10^4$ & 0.048 & 0.70 & 1.84 & 5.77 & 0.32 & 1.9 & 0.53 \\
    -2.0 & 0.15 & $2.0  \pm 1.0$ & 0.25 & 0.47 & 2100 & 0.29 & 6.0 & 175 & 230 & 1.7 & $6.9\times10^4$ & 0.034 & 0.40 & 1.44 & 2.86 & 0.54 & 3.2 & 0.45 \\
    -1.2 & 0.15 & $3.8  \pm 1.1$ & 0.36 & 0.52 & 1800 & 0.31 & 6.0 & 190 & 210 & 1.6 & $7.7\times10^4$ & 0.032 & 0.36 & 1.39 & 2.60 & 0.60 & 3.6 & 0.44 \\
     0.0 & 0.15 & $6.5  \pm 0.7$ & 0.46 & 0.68 & 1300 & 0.35 & 6.0 & 185 & 200 & 1.3 & $8.9\times10^4$ & 0.026 & 0.25 & 1.26 & 1.96 & 0.57 & 3.4 & 0.55 \\
     2.0 & 0.15 & $3.6  \pm 1.0$ & 0.36 & 0.53 &  900 & 0.24 & 6.0 & 155 & 180 & 1.2 & $7.4\times10^4$ & 0.024 & 0.21 & 1.22 & 1.80 & 0.53 & 3.2 & 0.41 \\
     3.3 & 0.15 & $2.3  \pm 0.9$ & 0.21 & 0.43 &  720 & 0.20 & 6.0 & 140 & 180 & 1.0 & $5.1\times10^4$ & 0.020 & 0.15 & 1.16 & 1.53 & 0.62 & 3.7 & 0.30 \\
    \hline
    \end{tabular}
\end{table*}

For clarity, we have presented results in Fig.~\ref{fig:beta} with only
experimental uncertainty. There are also uncertainties associated with the
turbulence model. Changes in the results due to most of these uncertainties are
primarily expected to be quantitative, with the picture of reduced mean density
remaining. One uncertainty comes from the possibly non-equilibrium nature of any
turbulence at stagnation. The turbulent velocity decreases in time during
stagnation, as evidenced by the decreasing non-thermal energy excess per-ion
($\Tieff - T_i$) in Table~\ref{tbl:data}. However, contrary to the turbulence
simulations usually considered for modeling (e.g. in \citep{hopkins:2013a}), the
total mass is not constant in time: at least initially, plasma continues to flow
into the stagnation region. Using the isothermal turbulence $r_{pl}$ and $n_e$
in Table~\ref{tbl:data}, $r_\mathrm{pl}^\mathrm{turb}$, $n_e^\mathrm{turb}$,
along with a turbulent energy per particle of $\Tieff - T_i$, yields a total
turbulent energy in the stagnation region that remains relatively constant from
$t = -3.4$ ns to $t=0$ ns, then falls. Notably, the timescale for this observed
fall (a few ns) is the dynamical timescale expected for supersonic turbulence
\citep{maclow:1998a,maclow:1999a} with these flow speeds and length scales;
importantly, it is much faster than the viscous timescale, without a cascade.
Although the present density PDF model works in a variety of cases, it has
typically been tested in situations with equilibrium forcing, which may not be
the best analog for the present case.

Assuming the model applies, there are still uncertainties. One is the degree to
which turbulence in this stagnating plasma would be isothermal. Conduction and
turbulent timescales are not well-separated, so that an accurate determination
of the degree of isothermality would likely require detailed simulations, as in
other topic areas~\citep{pavlovski:2002a, pavlovski:2006a}. The turbulence model
used here applies in the non-isothermal case, with different $\theta$,
$\sigma_{s,V}^2$~\citep{federrath:2015a}. At this level, non-isothermality is
expected to only modestly change the parameters in Table \ref{tbl:data}, 
although then the inferred $T_e$ and $M$ will also need to be reconsidered,
because non-isothermality would have a pronounced effect on the local plasma
emissivity (it depends rather strongly on $T_e$), requiring modifications to
Eqs.~(\ref{eq:I}) and (\ref{eq:R})---and therefore  also to (\ref{eq:beta2}) and
(\ref{eq:beta1}).

Any magnetic fields in the stagnating region could alter
the values of $\theta$ and $\sigma_{s,V}^2$ \citep{padoan:2011a, molina:2012a,
hopkins:2013a}, although the form of the PDF remains valid. These corrections
should be small because the plasma pressure is much higher than the
magnetic pressure in the stagnation region ($\beta_{\rm{magnetic}}\gtrsim
20$)~\cite{rosenzweig:2014a}.

Even within the isothermal turbulence PDF model, there are uncertainties.
Simulations show substantial spread in values of the PDF parameters around the
expressions for $\theta$ and $\sigma_{s,V}^2$, see, e.g.,~\cite{hopkins:2013a}.
Apart from modeling errors, spread in these values can be physical, due to the
fluctuations of turbulence \citep{lemaster:2008a}. The correct value of $b$,
presently taken to be $b=0.4$, is uncertain. Generally, $b \in
\left[1/3,1\right]$~\citep{federrath:2008a,federrath:2010a}, with $b = 1/3$
occurring for solenoidal (divergence free) forcing~\citep{federrath:2008a} and
$b = 1$ occuring for compressive (curl-free) forcing. Equal parts solenoidal and
compressive forcing gives $b \approx 0.4$~\citep{federrath:2010a}. For a
z-pinch, one might expect largely compressive forcing. Given this
uncertainty, one could calculate the range of $\beta$ in Fig.~\ref{fig:beta}a
including also uncertainty in $b$. A larger $b$ yields a lower range for
$\beta$, while a smaller $b$ yields a higher range.

The ion temperatures in Table~\ref{tbl:data} are inferred through a calculation
involving the electron--ion temperature equilibration time \citep{kroupp:2011a}.
Since this time is density dependent, it will be affected by density
fluctuations. The equilibration timescale is faster in the high density regions,
which dominate the measurements, thus, the ion temperature may be driven
slightly closer to the electron temperature. Since the electron--ion temperature
equilibration timescale is already very fast, this is expected to cause $T_i$ to
be a few percent lower than in Ref.~\cite{kroupp:2011a}.

The underlying atomic model used for the present analysis is the same as in the
previous study~\cite{kroupp:2011a} and, therefore, no additional uncertainties
have been introduced. In fact, the associated inaccuracy may be
surprisingly low, as the Monte-Carlo analysis of uncertainty propagation in
collisional-radiative models indicates~\cite{ralchenko:2016a}.
So far, we have neglected possible opacity effects. Fortunately, the satellites
used have a negligible optical thickness. The bound--free and free--free
(bremsstrahlung) radiation that contributes to the PCD signal is also optically
thin, however strong bound--bound transitions are not. This requires a
modification of Eq.~(\ref{eq:I}) which cannot be represented analytically.
However, the plasma absorption coefficients, similar to the emission ones, for
these transitions scale as $n_e^2$. Therefore, the difference from the
uniform-plasma model (in which the opacity was properly accounted for
numerically) should vanish in the lowest order.

The mechanism generating the (non-radial) hydrodynamic motion is unclear; while
energy is dumped in the hydrodynamic motion in the process of
stagnation~\citep{maron:2013a}, this hydrodynamic motion could be seeded by
turbulence generated and carried along during the compression itself, or could
be generated entirely at stagnation. In either event, there are important
implications, both for z-pinches, and more broadly.

If the (turbulent) hydromotion is generated and carried along during the
compression, these z-pinches represent a test bed for the properties of plasma
turbulence undergoing compression. These properties are relevant for a proposed
novel fast ignition or x-ray burst generation scheme~\citep{davidovits:2016a,
davidovits:2016b}. Of particular interest is that the present hydromotion is
supersonic, the regime in which these schemes would operate. Further, the
behavior of compressing supersonic turbulence is of critical interest in
astrophysics, particularly for molecular cloud dynamics~\citep{robertson:2012a,
davidovits:2017a}. Supersonic turbulence behavior has been related to the star
formation efficiency~\citep{elmegreen:2008a}, the core mass/stellar initial mass
functions~\citep{padoan:2002a, ballesteros:2006a, hennenbelle:2008a}, and
Larson's laws~\citep{kritsuk:2013a}.

If the hydrodynamic motion is generated at stagnation, and then
decays, its properties could still be of astrophysical interest (see, e.g.,
\citep{maclow:1998a, maclow:1999a, smith:2000a, ostriker:2001a, pavlovski:2002a,
pavlovski:2006a, lemaster:2008a, kitsionas:2009a, davidovits:2017a}). To the
extent generation and/or decay of the hydrodynamic motion at stagnation can be
observed, studies of supersonic turbulence in z-pinches could serve as a new and
important area for laboratory astrophysics. Indeed, in the present study, not
all values of the turbulent PDF parameters, $\theta,\sigma_{s,V}^2$, will be
consistent with the observations; more measurements could help to constrain
turbulent properties. Z-pinches such as the present may yield other cross-over
opportunities with astrophysics, for example, in turbulent density PDF
measurement techniques (e.g.~\cite{ostriker:2001a, brunt:2010a}), or in
mechanisms for turbulent generation and forcing in complex plasma environments
(e.g.~\cite{gritschneder:2009a}).

The present analysis is likely relevant to high-current implosions, like z-pinch
experiments on the Z machine~\cite{jones:2006a}. Indeed, based on the plasma
parameters given in~\cite{maron:2013a}, Re is also high ($\sim \! 10^4$), and M
is similar to the case analyzed here. This analysis may also be relevant in
inertial confinement experiments that observe large quantities of residual
hydrodynamic motion.

Even though the present experiments are very well diagnosed by z-pinch
standards, one should consider additional measurements for verifying the picture
of a turbulent stagnation. To this end, other spectroscopy methods can be
useful. For example, one can try to study the density by the use of the Stark
broadening of high-$n$ transitions in hydrogen-like Ne or lower-$Z$ species (C,
N, or O) that can be mixed with the puffed neon.

In summary, a new analysis of stagnating pinch data, replacing the assumption of
uniform plasma with density variations consistent with a turbulent plasma,
advances a picture of supersonically turbulent stagnating plasma. This picture
is not only consistent with the observations, it improves the agreement with
them. The mean plasma density is reduced by a factor $\sim$2. While there is
uncertainty in the precise value of this reduction, the general picture, of a
data analysis in the presence of highly turbulent stagnating plasma reducing the
inferred stagnation density compared to the uniform case, is believed to be
robust and widely relevant. Beyond aiding our understanding of z-pinches, we
hope this study has shown fertile ground for relation to problems of
astrophysical interest.

\begin{acknowledgments}
Y.M. is grateful to M. Herrmann, A. L. Velikovich, and E. P. Yu for enlightening
discussions. This work was supported by BSF--NSF (USA) and BSF 2014714. The work
of E.K., E.S., A.S., V.I.F., and Y.M. was supported in part by the Israel
Science Foundation and the DOE--Cornell University Excellence Center (USA). The
work of S.D. and N.J.F. was supported by NNSA 67350-9960 (Prime \# DOE
DE-NA0001836) and by NSF Contract No. PHY-1506122.
\end{acknowledgments}

\end{document}